\documentclass[aps,prl,preprint,superscriptaddress]{revtex4}
\usepackage{graphicx,subfigure}

\begin{document}

\title{Mesoscopics and the High $T_c$ Problem}


\author{D.J. Scalapino}
\address{Department of Physics, University of California,\\
Santa Barbara, CA 93106-9530, USA}
\author{E. Berg}
\address{Department of Physics, Harvard University, Cambridge, MA 02138, USA}
\author{S.A. Kivelson}
\address{Department of Physics, Stanford University, Stanford, CA 94305-4045, USA}


\begin{abstract}
Mesoscopic physics concerns itself with systems which are intermediate
between a single atom and a bulk solid. Besides the many intrinsically
interesting properties of mesoscopic systems, they can also provide
physical insight into the physics of bulk systems. Here we discuss three
examples of this from the field of high temperature superconductivity.
\end{abstract}

\maketitle

Many interesting electronic materials, especially those that exhibit high
temperature superconductivity, lie in an intermediate coupling regime where
the strength of the interaction is comparable to the electron bandwidth. The
absence of any small parameter makes analytic approaches to such problems
difficult. Conversely, in the absence of a small parameter there is every
reason to expect the correlation lengths associated with any form of electronic
order to be of order 1 in units of the lattice constant (barring an accident
which places the system close to a critical point). Thus, with some sensible
analysis (and with a little bit of luck), the properties of the system in the
thermodynamic limit may be apparent in the properties of mesoscopic systems,
even systems that are small enough that they can be studied by essentially
exact numerical methods. Here we see how three such studies have provided
insight into the high $T_c$ problem.

Following the discovery of the high $T_c$ cuprates there were various
suggestions of ways in which the strongly repulsive Hubbard interaction, $U$,
between two electrons on the same site could (paradoxically) produce high
temperature pairing in a doped antiferromagnetic insulator. Numerous studies
began with the  $t-J$ model on a square lattice, which can be thought of as
the strong coupling limit of the Hubbard model. Here it is assumed that $U$
is sufficiently large to prevent double occupancy of any site, leaving
a one-electron near-neighbor hopping term $t$ and an exchange coupling $J$.
An early argument for pairing was based upon the observation that if one adds
two holes to the half-filled (one-electron per site) system, then eight
exchange $J$ bonds are broken if the holes were well separated. However,
if the two holes are placed on near neighbor sites, only seven exchange $J$
bonds are broken. Thus there is an effective near-neighbor attraction between
the holes. This picture however was soon seen as more applicable to the phase
separation regime \cite{ref:1}.  Alternatively, in the context of the
resonance-valence-bond \cite{ref:A} approach, variational calculations using
Gutzwiller projected  wavefunctions \cite{ref:G} and auxiliary-boson
meanfield \cite{ref:Ru,ref:Ko} calculations found a superconducting state
with $d$-wave symmetry in the $t-J$ model. From a more weak-coupling perspective,
the idea of spin-fluctuation exchange mediated pairing near an antiferromagnetic
instability \cite{ref:B} was also found to lead to $d$-wave pairing due to the
increasingly positive strength of the pairing interaction at large momentum
transfer. However, none of these approaches gave a simple, crisp real space
picture, especially one that makes clear why $d$-wave rather than extended
$s$-wave symmetry is preferred.

To address this, Trugman and one of the authors \cite{ref:2} decided to imagine
that a 4-site plaquette was extracted from the lattice. The ``undoped" groundstate
of the $t-J$ model on a plaquette with 4 electrons is a singlet having a wave function
\begin{equation}
  \vert\psi_0(4)\rangle=(\Delta^+_{12}\Delta^+_{34}-\Delta^+_{23}\Delta^+_{41})\
    \vert0\rangle.
    \label{eq:1}
\end{equation}
Here, $\Delta^+_{ij}=(c^+_{i\uparrow}c^+_{j\downarrow}-c^+_{i\downarrow}c^+_{j\uparrow})/\sqrt2$
creates a singlet pair on sites $ij$ and we have numbered the sites of the
plaquette in a clockwise manner. This state is odd under a $\pi/2$ rotation.
The two-electron groundstate
\begin{equation}
  \vert\psi_0(2)\rangle=N(c^+_{2\downarrow}c^+_{1\uparrow}+c^+_{4\downarrow}c^+_{1\uparrow}+\cdots)\
    \vert0\rangle
    \label{eq:2}
\end{equation}
has spin 0 and is invariant under a $\pi/2$ rotation. Therefore the pairfield
annihilation operator that connects the zero-hole (4-electron) and two-hole
(2-electron) groundstates of the $2\times2$ plaquette must transform as
$d_{x^2-y^2}$. The same calculation can be  performed for the Hubbard model on
a single plaquette; while the wave-functions are somewhat more complex, in
this case, the symmetry of the 2 and 4 electron ground-states are invariant
for any $U$ in the range $0 < U <\infty$. As Carlson et al. \cite{ref:3} noted,
it showed the robustness of the $d$-wave character of the pairing in $t-J$ and
Hubbard models.

Of course studies of a 4-site model could not say anything about the
possibility of superconducting order. However, it turned out that studies of
2-leg $t-J$ and Hubbard ladders yielded important insights concerning the
character of the superconducting groundstate. The study of 2-leg ladders
was motivated by a simple picture based upon the case in which the rung exchange
interaction $J_r$ is large compared to the near neighbor leg exchange $J_\ell$.
In this limit, for the undoped Heisenberg ladder, spin singlets tend to form on
the rungs leading to a spin gapped groundstate. Then when holes are added,
where $J_r > t$, they would occupy sites on either side of a rung so
as to break only one exchange rung coupling. A measure of the spatial correlation
of these rung hole pairs would then allow one to probe the superconducting order.
Based on this large $J_r/J_\ell$ picture, it was initially a surprise when
numerical calculations \cite{ref:4} showed that at half-filling the spin gap
persisted to small values of $J_{r}/J_{\ell}$.  In addition, for the hole doped
ladder, despite the fact that $t > J_r$, the equal time pairfield-pairfield
correlations appeared to have a power law decay, indicative of quasi-long-range
``superconductivity". In later work \cite{chakravarty,ref:5}, it was understood
that the ladder would have a spin gap at half-filling for any finite
$J_{r}/J_{\ell}>0$, and that the groundstate of the doped ladder is a
Luther-Emery \cite{ref:13} liquid.  Furthermore, the pair structure is
$d$-wave-like in the sense that the rung and leg pairfield amplitudes have
opposite signs. We now also know that, in the limit as  the length of the
ladder tends to infinity, the $t-J$ ladder has perfect Andreev reflection in
response to an externally applied pairfield at one end of the ladder\cite{ref:6}.
The 2-leg $t-J$ and Hubbard ladders now represent some of the best understood
models of strongly correlated electron systems.

Admittedly, since the plaquette and the ladder are, respectively, zero and one
dimensional systems, neither can support a superconducting phase with a finite
transition temperature.  However, in many cases it is possible to analyze the
phase diagram of a higher dimensional system constructed as an array of
{\em weakly} coupled mesoscale structures, starting from the exact numerical
solution of the isolated structure, and treating the coupling between clusters
in the context of perturbation theory \cite{imry}.  Studies of arrays of weakly
coupled two-leg ladders\cite{tsun,arrigoni} and plaquettes\cite{assa,hong} (the
``checkerboard Hubbard model'') lead to rather complex phase diagrams with many
competing phases, even where the above analysis shows strong superconducting
correlations on the isolated cluster.  Nonetheless, among those phases there
are robust regions of $d$-wave, or $d$-wave-like superconductivity.

\begin{figure}
\includegraphics[width=0.7\textwidth]{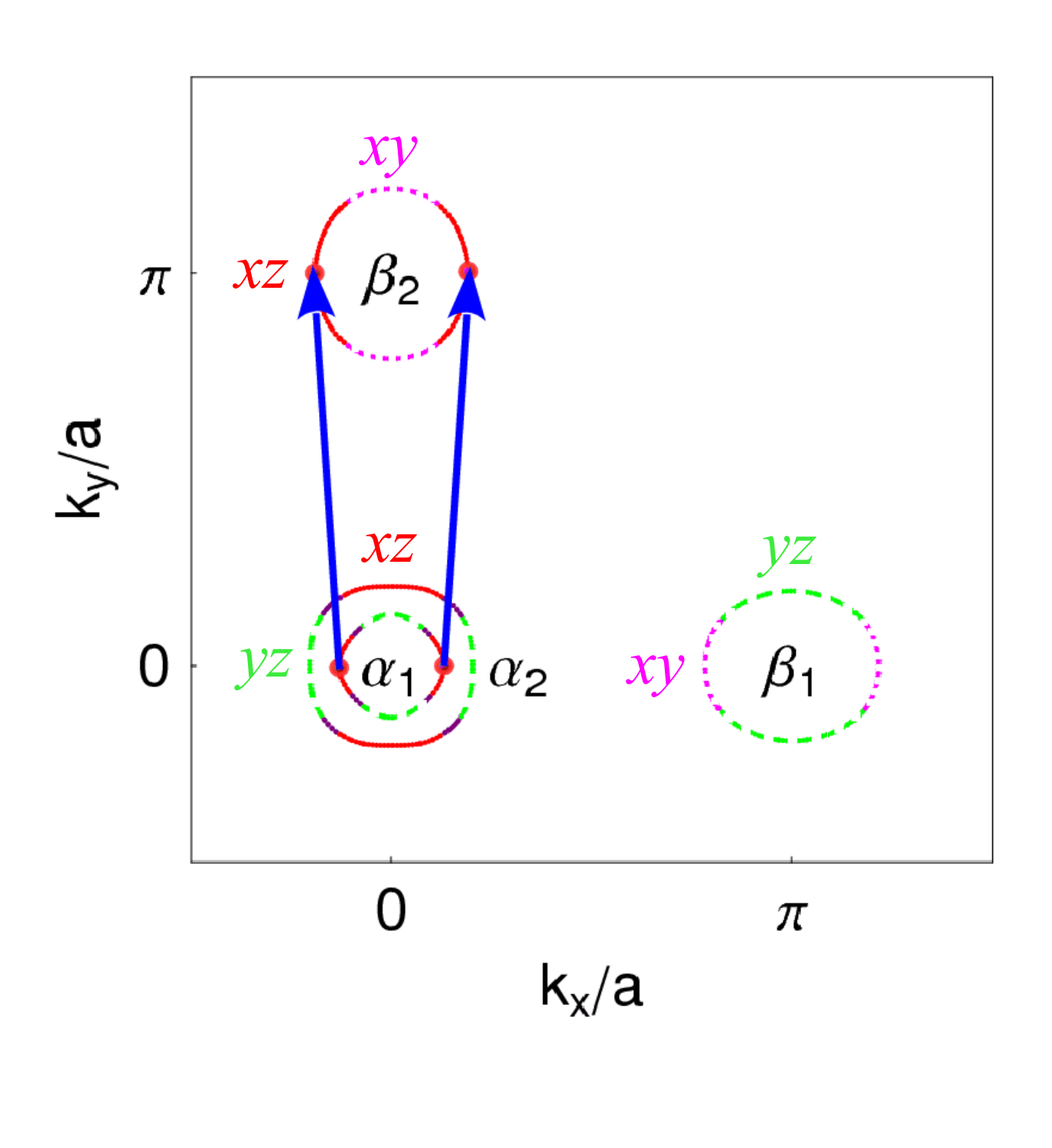}
\centering \caption{The Fermi surfaces for a five-orbital tight
binding model of the Fe-pnictides. The main orbital contributions
to the Bloch states are indicated: $d_{xz}$ (solid line), $d_{yz}$
(dashed line) and $d_{xy}$ (dotted line). The arrows illustrate
the type of $d_{xz}-d_{xz}$ inter-Fermi surface scattering
processes that lead to pairing in the spin-fluctuation-exchange
calculations.} \label{fig:1}
\end{figure}

As a final example of insights gained from studies of small systems, we turn to
calculations on a 2-leg ladder model of an Fe-pnictide superconductor \cite{ref:7}.
Figure~\ref{fig:1} shows the typical Fermi surfaces of the Fe-pnictide materials
in an unfolded (1 Fe/cell) Brillouin zone. There are two-hole Fermi surfaces
$\alpha_1$ and $\alpha_2$ around the $\Gamma$ point and two-electron Fermi
surfaces $\beta_1$ and $\beta_2$ around $(\pi,0)$ and $(0,\pi)$. The symbols
indicate the dominant $d$-orbital contributing to the Bloch state on the
indicated portion of the Fermi surfaces. In weak coupling, RPA \cite{ref:K,ref:10}
and functional renormalization group \cite{DHLee} calculations suggest that the
pairing arises from the scattering of time-reversed-pairs from the
$d_{xz}$-dominated states on the $\alpha_1$ Fermi surface to paired states
with the same orbital character on the $\beta_2$ Fermi surface, and from the
analogous processes involving pairs in the $d_{yz}$ dominated states on the
$\alpha_1$ and $\beta_1$ Fermi surfaces. This is illustrated in Fig.~\ref{fig:1}
for the $d_{xz}-d_{xz}$ pair scattering.

\begin{figure}
\includegraphics[width=0.7\textwidth]{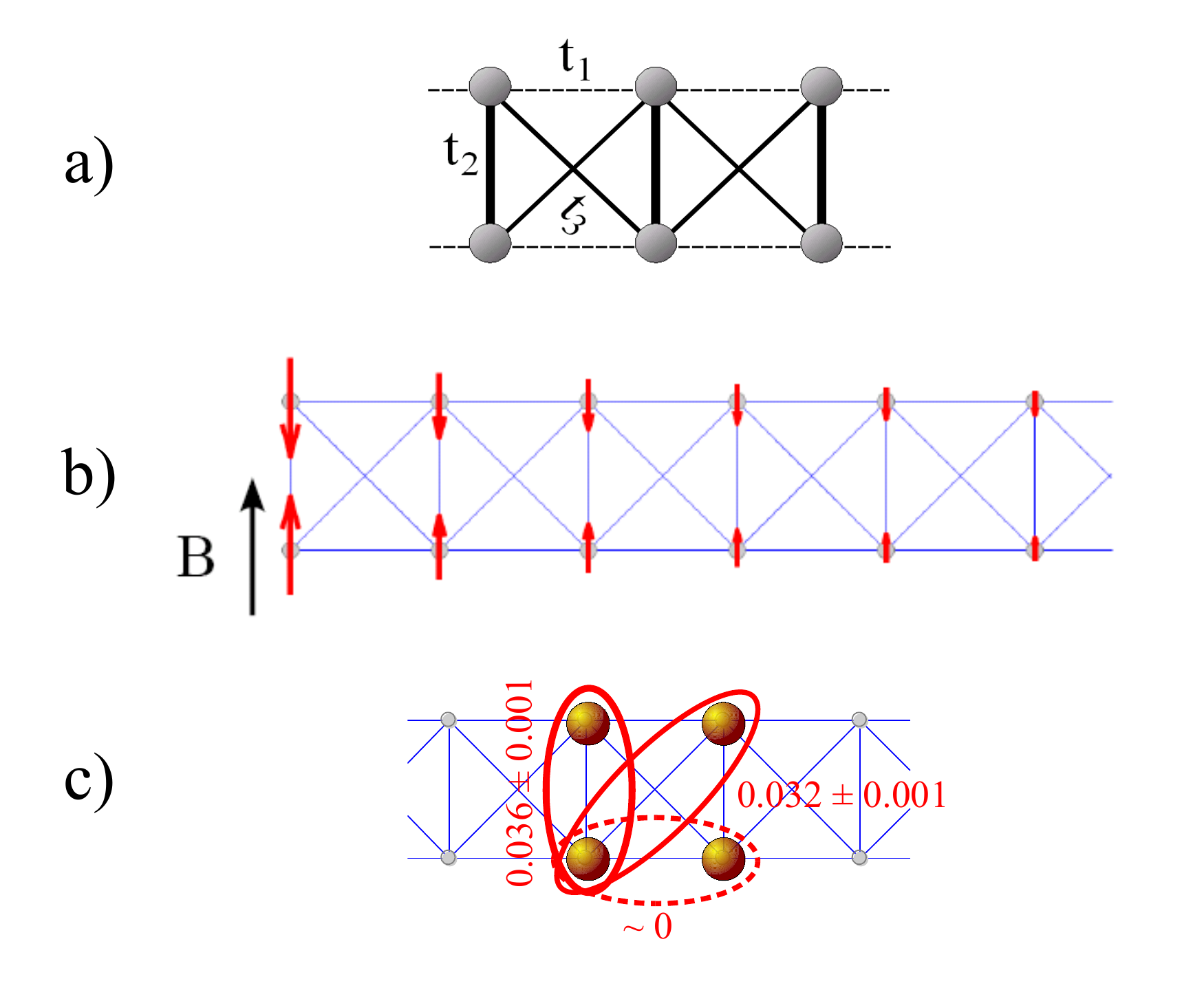}
\centering \caption{a) An Fe two-leg ladder with $t_1=-0.32$,
$t_3=-0.57$ and $U=3$ in units where $t_2=1$. These hopping
parameters were chosen to fit the DFT calculation of the
bandstructure for cuts with $k_y=0$ and $\pi$.; b) The spin
structure $\langle S^z(\ell_x,\ell_y)\rangle$ induced on the
undoped Fe ladder when an external magnetic field is applied to
the lower left hand site.; c) The singlet pairfield
$\langle\Delta_{ij}\rangle$ induced across a rung, along a
diagonal and along a leg at a distance 10 sites removed from the
end of a $32\times2$ Fe-ladder with a unit external pairfield
applied to its end rung.} \label{fig:2}
\end{figure}

In order to use numerical methods to study these processes in the intermediate
to strong coupling limit, the problem needs to be simplified. If we accept that
the type of scattering processes shown in Fig.~\ref{fig:1} capture the essential
physics, we can focus exclusively on pair scattering involving two bands and
only one orbital. The resulting two-leg Hubbard ladder retains the $d_{xz}$
states along two cuts through the 2d BZ, ${\bf k}=(k_x,0)$ which passes through
the $\alpha_1$ Fermi surface, and ${\bf k}=(k_x,\pi)$ which passes through $\beta_2$.
This reduces the problem to that of the two-leg Hubbard ladder shown in
Fig.~\ref{fig:2}a which can then be studied using the numerical density matrix
renormalization group (DMRG) \cite{ref:D}.
The one-electron hopping parameters $t_1=-0.32$, $t_3=-0.57$ in
units of $t_2$, were chosen to reproduce the density functional
bandstructure \cite{ref:8} near the Fermi surface for $k_y=0$ and
$\pi$.  The repulsion $U$ between two electrons in the same
orbital was varied in the range 3--4 in units of $t_2$.

In the undoped, one electron per site, limit one finds the expected spin gapped
groundstate. By applying a magnetic field to one of the end sites of the ladder,
the resulting expectation value of the spin appears as shown in
Fig.~\ref{fig:2}b. Here one sees ``stripe"-like $(0,\pi)$ spin correlations
which decay with a slow exponential due to the spin gap. Hole doping the system
and applying an external pairfield on the end rung of the ladder, one obtains
the pairfield singlet amplitudes illustrated in Fig.~\ref{fig:2}c. Here a
pairfield of unit strength was applied to the left end rung and
Fig.~\ref{fig:2}c shows the strength of the induced pairfields
$\langle\Delta_{ij}\rangle$ ten sites to the right. The relative positive sign
of the pairfield on the rung and diagonal and the negligible value of the
pairfield on the legs is expected if the gap changes sign between the $\alpha_1$
and $\beta_2$ Fermi surfaces \cite{ref:9,ref:K,ref:10}.

\begin{figure}
\includegraphics[width=0.7\textwidth]{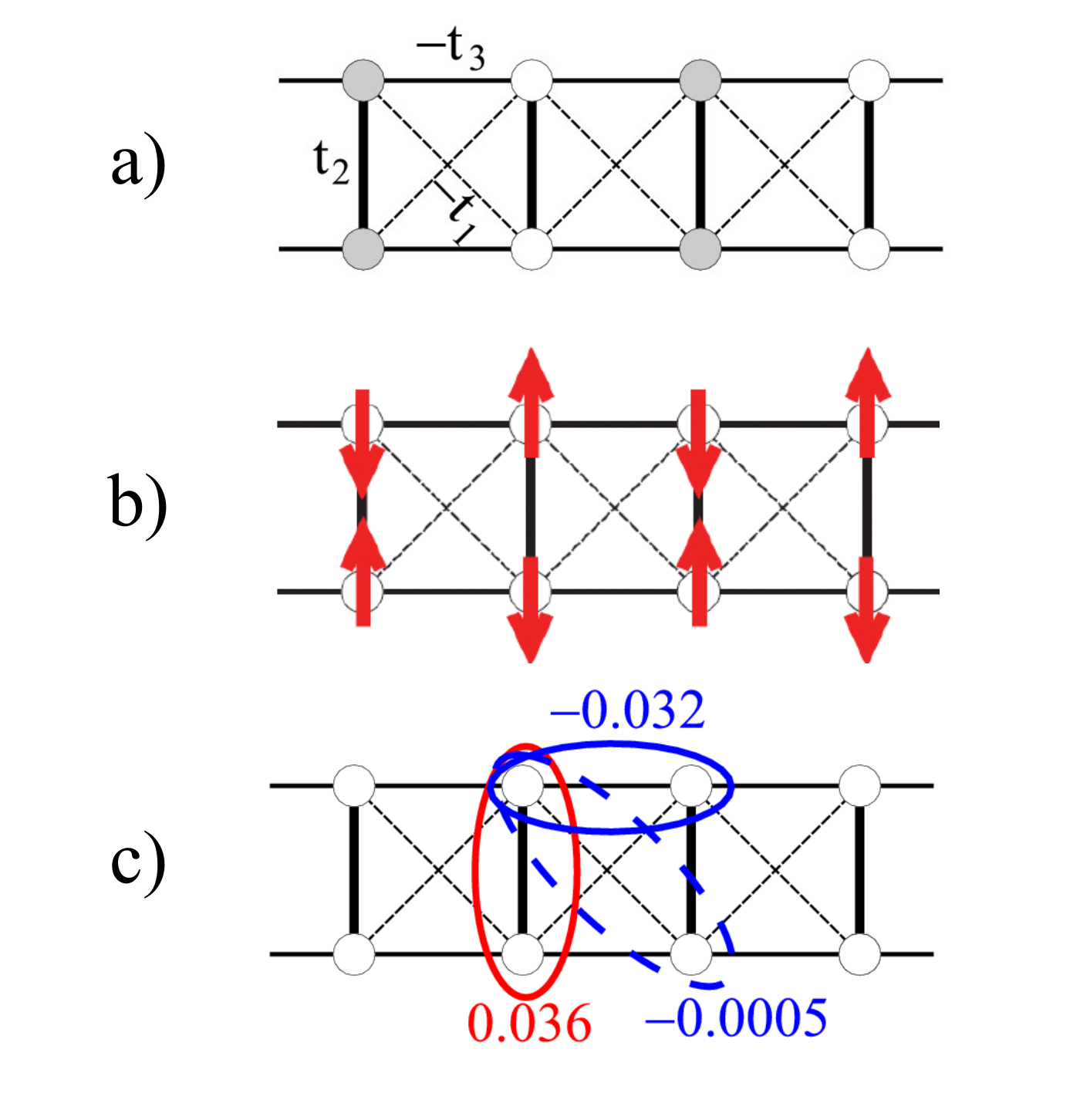}
\centering \caption{a) Here every other rung of the ladder shown
in Fig.~\protect\ref{fig:2}a has been twisted by $180^\circ$ and
the phases of the orbitals denoted by the shaded sites have been
changed by $e^{i\pi}=-1$.; b) The spin expectation values of
Fig.~\protect\ref{fig:2}b for the twisted ladder show the spin
gapped $(\pi,\pi)$ antiferromagnetic behavior of the familiar
cuprate ladder.; c) The induced pairfield correlations of
Fig.~\protect\ref{fig:2}c become the familiar $d$-wave-like
pairing correlations seen for models of the cuprate ladders.}
\label{fig:3}
\end{figure}

The stripe-like SDW pattern of the spin correlations in the undoped system as
well as the structure of the pairfield are consistent with what is found in
the RPA calculations \cite{ref:K,ref:10}. However, what we found most
interesting was the relationship between the Fe-ladder and the previously
studied 2-leg cuprate ladder. This is illustrated in Fig.~\ref{fig:3}.
Here in Fig.~\ref{fig:3}a, every other rung has been twisted by $180^\circ$ and
the phase of the $d_{xz}$-orbit has been changed by $\pi$ on each of the sites of the
twisted rung. In this way, the rung one-electron hopping matrix element remains
$t_2$, but the leg and diagonal hoppings are changed to $-t_3$ and $-t_1$,
respectively. The dominant hoppings on the twisted Fe-ladder are along the legs
and rungs with only a weak diagonal hopping. The spin correlations shown in
Fig.~\ref{fig:3}b, obtained by twisting every other rung of Fig.~\ref{fig:2}b,
look just like the spin gapped AF correlations of the previously studied 2-leg
Hubbard cuprate ladder. Because of the twist and the phase change $e^{i\pi}=-1$
of the orbitals on the sites of the twisted rungs, the pairfield correlations
take on the $d$-wave-like form shown in Fig.~\ref{fig:3}c. In short, the twist
maps $(\pi,0)$ magnetic and sign-changing $s$-wave pairing correlations on the
Fe-ladder into $(\pi,\pi)$ magnetic and $d$-wave-like pairing correlations in the
cuprate ladder! Finally, it turns out that the ratio of the leg-to-rung hopping
0.57 obtained from the fit to the Fe-pnictide DFT bandstructure is near the
value which was previously found \cite{ref:11} to give the slowest pairfield
decay for a cuprate ladder. Thus this Fe-ladder turns out to simply be a
twisted version of the cuprate 2-leg Hubbard ladder with parameters near those
which are optimal for pairing.  This provides a direct link between the physics
of these two materials.

Now, as noted by Joe in his book {\it Introduction to Mesoscopic Physics}
\cite{ref:12}, ``the interest in studying systems in the intermediate size
range between microscopic and macroscopic is not only in order to understand
the macroscopic limit. Many novel phenomena exist that are intrinsic to
mesoscopic systems." Here we have only touched on some examples where strongly
correlated mesoscopic models have been introduced in the hope that they can
provide some insight into the macroscopic high $T_c$ problem. It is natural
to ask whether there aren't novel mesoscopic phenomena as well. Indeed,
there are. For example, the difference between the even- and odd-legged
Heisenberg ladders in which the even-leg ladders have a spin gap while the
odd-leg ladders are gapless is a mesoscopic width effect \cite{tsun}.  It
is also known that while the doped 2-leg ladder goes into a Luther-Emery phase
\cite{ref:13}, it takes a finite doping to bring the 3-leg ladder into this
phase \cite{ref:R,ref:W}. Ladders also appear in the striped phase of the
cuprates and, a better understanding of the mesoscopic properties of multi-leg
ladders may shed light on the recently proposed $\pi$-phase shifted $d$-wave
stripes \cite{ref:14}.

\section*{Acknowledgments}

It is a pleasure to contribute to this volume in honor of Yoseph Imry's 70th
birthday. We know Joe as a scientist, teacher, co-worker and friend.  DJS met
Joe in the early seventies when Joe first came to UCSB. He remembers how they
often talked about how one might gain insight into the basic physics of a
macroscopic system from calculations on small subsystems and how these
subsystems had their own interesting features. All of us are eager to recognize
how much we have learned from Joe's work, over the years concerning the
intrinsic, subtle and beautiful physics one finds in the mesoscopic world.
This work was supported in part by the National Science Foundation under the
grant PHY05-51164 at the KITP.  DJS acknowledges the Center for Nanophase
Materials Science at ORNL, which is sponsored by the Division of Scientific User
Facilities, U.S.\ DOE. EB was supported by the NSF under grants
DMR-0705472 and DMR-0757145 at Harvard. SAK was supported, in part, by the NSF
grant number DMR-0758356 at Stanford.

\end{document}